\documentclass[aps,twocolumn,epsfig,graphics,showpacs,floatfix,mathbbm]{revtex4}
\usepackage{amsmath,amsfonts,amssymb,graphics,graphicx,epsfig,color,times,bbm}

\begin{document}
    
\bibliographystyle{apsrev}
\newcommand{\proofend}{\hfill\fbox\\\medskip }
\newcommand{\rr}{{\mathbbm{R}}}
\title{Exact decoherence to pointer states in free open quantum systems is universal}

\author{
Jens Eisert}

\affiliation{
Institut f{\"u}r Physik, Universit{\"a}t Potsdam,
Am Neuen Palais 10, D-14469 Potsdam, Germany\\
QOLS, Blackett Laboratory, Imperial College London,
Prince Consort Road, London SW7 2BW, UK
 }

\date{\today}

\begin{abstract}
In this letter it is shown that exact decoherence to minimal 
uncertainty Gaussian pointer states is generic for free quantum particles
coupled to a heat bath. More specifically, the paper is concerned with
damped free particles 
linearly coupled under product initial conditions 
to a heat bath at arbitrary temperature, with 
arbitrary coupling strength and spectral densities covering the 
Ohmic, subohmic, and supraohmic regime.
Then it is true that there exists a time 
$t_{c}$ such that for times 
$t>t_{c}$ the state can always be exactly represented as a mixture (convex 
combination) of particular
minimal uncertainty Gaussian states, regardless of and independent from
the initial state. This exact `localisation'
is hence not a feature specific to high temperatures and weak damping 
limit, but is rather a generic property of damped free particles.
\end{abstract}


\maketitle


There is long tradition of approaching 
the questions of how and to what extent 
classical properties of quantum systems
emerge dynamically due to
the unavoidable coupling to their environment.
Essentially any quantum system interacts
to some extent with other external degrees of freedom, 
which in turn may be said to 
monitor certain properties
of the quantum systems 
\cite{Dec2,Caldeira,Halliwell,Strunz,Eisert,Coh,Kiefer,Knight,Exact,Exp}. 
This yields decoherence, which results in a loss of purity of
initially pure states of a distinguished quantum system
coupled to an environment. Not all initial quantum 
states in such a dilation are
yet equally `fragile' to this interaction: there is a
small set of initial states that is often relatively
robust with respect to this interaction. The term
pointer states has been coined for such states,
owing the name to 
models for quantum measurement where
the pointer basis is essentially determined by
the interaction of the apparatus with the
external degrees of freedom \cite{Dec2}.

For harmonic and free quantum systems linearly
coupled to a heat bath consisting of harmonic systems
this general mechanism is very well-understood indeed.
For example, if one prepares a single mode in a pure state
in order to let it very weakly interact  
with an environment in the Gibbs state 
corresponding to a very high temperature, 
which one is the state that produces
the least entropy over one cycle of the oscillator? 
In retrospect it hardly comes as a surprise that this
is a coherent state, a minimal uncertainty Gaussian 
state \cite{Coh}. 
Most attention has probably 
been devoted to thoroughly
understanding the dynamics of harmonic
and free quantum systems in this limit of weak coupling and
high temperatures \cite{Dec2,Caldeira,Halliwell,Coh}. 
In this limit in particular, decoherence time scales have been identified
\cite{Dec2}.
But also exact quantum master equations, generators of dynamical
maps, have been derived and scrutinized 
in great detail \cite{Exact,Hu,Hakim,Previous,BroMo}. 
After all, the dynamics of open harmonic or free
quantum systems can not be described other than being 
well-understood. What else is there to ask for?

A question that seems to have been overlooked so far, yet, is
the following: To what extent is exact decoherence in 
free quantum systems to pointer states actually generic?
This question is most appealing in case of the free
damped quantum particle \cite{Hakim,BroMo}, 
where there is no equilibrium Gibbs state.
More specifically: is it true that starting from an arbitrary 
quantum state, after a fixed finite time $t_{c}$
(independent of the initial state), the state of the 
system is exactly indistinguishable from a mixture, a convex combination, 
of minimal uncertainty Gaussian states for all times $t>t_{c}$? 
In this sense the free quantum system may be said to be in a
situation that can operationally not 
be distinguished locally 
from the following situation: 
the particle is somewhere, in a 
minimal uncertainty Gaussian state, one simply does not 
know where in phase space. 
That this is the case seems fairly
plausible for the case of high temperatures and weak damping.
A significant first step in this direction has indeed
been achieved very recently
by Diosi and 
Kiefer in Ref.\ \cite{Kiefer}, showing that this intuition is 
indeed correct for the approximate generator for the dynamical
map in the limit of negligible friction and at high temperatures.
But is this a generic feature of free quantum 
systems that are
linearly coupled to an environment in a dilation, 
and true not only for specific regimes, 
but 
for any coupling strength,
any non-zero temperature, and Ohmic, subohmic, as well as 
supraohmic damping? This is the question that will 
be addressed (and answered) 
in this paper.

A free quantum system linearly
coupled to a heat bath of oscillators will be
investigated, where the distinguished system is 
initially in an arbitrary (and potentially very 'non-classical' state), 
whereas the environment is prepared in the Gibbs
state, which corresponds to an initial product state, such that
the time evolution of the state of the free quantum system 
amounts to a completely positive dilation \cite{NoteFact}.
No assumptions will be made concerning the
temperature of the environment and the strength of the coupling;
for the class of non-vanishing spectral densities
any $C^{\infty}$-function $I:\rr^{+}
\longrightarrow \rr^{+}$ could be allowed for
with 
\begin{equation}\label{spectraldensity}
	\lim_{\omega \rightarrow 0}
	I(\omega)/\omega^p=\zeta  >0
\end{equation}
for some $p\in(0,2)$. This will
be referred to as Ohmic damping 
when $p=1$, otherwise as subohmic (for $p<1$) or supraohmic
(for $p>1$). 
This is an already solved problem
in the sense that quantum master equations are known,
and hence, the argument draws heavily from known results
on generators of dynamical maps \cite{Hu,Ford},
and from earlier results on the long-time behavior  in
quantum Brownian motion \cite{BroMo}.
The starting point is the equation of motion of the
reduced density operators as derived in 
Ref.\ \cite{Hu}, in the integrated form as 
presented in the recent paper Ref.\ \cite{Ford}. Later, 
ideas will be used very similar to the ones in Ref.\ \cite{Kiefer}.

The equation of motion of the free particle
is for the subsequent purposes most
conveniently be expressed in phase space
in terms of the Wigner function 
$W:\rr^2 \times \rr^+\longrightarrow
\rr$ \cite{Wigner}, 
which is for each $t\in \rr^+$ 
the Fourier transform of the characteristic function,
dependent on $\xi=(\xi_1, \xi_2)\in \rr^2$, where
$\xi_1$ and $\xi_2$ correspond to 
position and momentum coordinates in phase space, 
respectively.
As a partial differential equation the Hu-Paz-Zhang equation
\cite{Hu}
reads \cite{Units}
\begin{eqnarray*}
	\partial_t W(\xi,t) && = - \xi_2 \partial_{\xi_1} W(\xi,t) + 
	\Omega^2(t) \xi_1 \partial_ {\xi_2} W(\xi,t)\\
	&&+  2 \Gamma(t) \partial_ {\xi_2} ( \xi_2 W(\xi,t)) \nonumber\\
	 &&+  \Gamma(t) h(t)
	\partial^2_{ \xi_2} W(\xi,t) + \Gamma(t) 
	f(t) \partial_{ \xi_1} \partial_{ \xi_2} 
	W(\xi,t),\nonumber
\end{eqnarray*}
where the $\Gamma,f,h,\Omega:\rr^{+}\longrightarrow\rr$ are time-dependent
coefficients for which explicit expressions are known.
The formal solution of this partial differential equation can be 
found for all system parameters \cite{Ford,Long}. 
The solution of the differential equation with
time-dependent coefficients as presented in Ref.\ 
\cite{Ford} is
given by
\begin{eqnarray}\label{solution}
	W(\xi,t)&=& \int d^2 \xi' \frac{1}{2\pi | M(t) |^{1/2}}
	\nonumber\\
	&\times &
	e^{ - (R(\xi,\xi',t) M(t)^{-1}  R(\xi,\xi',t)^T )/2}
	W( 	\xi' ,0)
	\nonumber\\
	R(\xi,\xi',t)&=&
	(\xi_1- \dot G(t) \xi_1'- G(t) \xi_2',
	\xi_2- \ddot G(t) \xi_1'- 
	\dot G(t) \xi_2'),\nonumber
\end{eqnarray}
where dots represent time derivatives.
Here, $G:\rr \longrightarrow
\rr$ is the Green's function, which
is $G(t)=0$ for $t<0$ and 
is for $t>0$ the solution
of the integral equation
\begin{eqnarray*}
    \ddot G(t) &+& \int_0^t ds \gamma(t-s) 
    \dot G(s) =0,
	\gamma(t)  = \int_0^\infty d\omega
	\frac{I(\omega)}{\omega} \cos(\omega t),
\end{eqnarray*}
with initial conditions $G(0)=0$, $\dot G(0)=1$,
in terms of the so-called damping kernel.
The $2\times 2$-matrix
\begin{equation*}
	M(t)=\left[
	\begin{array}{cc}
	A(t) & C(t)\\
	C(t) & B(t)\\
	\end{array}
	\right],
\end{equation*}
has coefficients that have in Ref.\ \cite{Ford} been 
expressed in terms
of correlation functions. 
On using the function $K:\rr^+\longrightarrow
\rr$, 
\begin{eqnarray*}
    K(t)=\frac{1}{\pi}
    \int_0^\infty
	d\omega
	\text{re}[\tilde \gamma(\omega+i 0^+	)] \, 
	\omega
	\coth(\beta \omega)\cos(\omega t),
\end{eqnarray*}
with $\tilde\gamma  = \int_0^\infty dt \gamma(t) e^{iz t}$,
the coefficients $A(t)$, $B(t)$, and $C(t)$
can be expressed as
\begin{eqnarray*}
	A(t) & =&
	\int_0^t ds \int_0^t ds'
	G(t-s)
	G(t-s') K(s-s'), \\
	B(t) &=&
	\int_0^t ds \int_0^t ds'
	\dot G(t-s)
	\dot G(t-s')
	K(s-s'),\\
	  C(t) &=&
	\int_0^t ds \int_0^t ds'
	G(t-s)
	\dot G(t-s')
	K(s-s'),
\end{eqnarray*}
as $G(0)=0$. Eq.\ (\ref{solution}), together with the subsequent
specifications forms the starting point of our analysis.

Eq.\ (\ref{solution})
can, using the transformation rule for multiple integrals,
be written in form of a product of
a time dependent determinant and a 
convolution with a Gaussian as 
\begin{eqnarray}\label{solution2}
	W(\xi,t)&=& \int d^2 \xi' \frac{1}{2\pi | M(t) |^{1/2}}
	e^{ - ((\xi-\xi') M(t)^{-1} (\xi-\xi')^T)/2}\nonumber\\
	&\times &
	\frac{1}{\left|V(t)
	\right|}
	W( V(t)^{-1}
	\xi' ,0)
\end{eqnarray}
where the $2\times 2$-matrix $V$ is given by
\begin{equation*}
	V(t)= 
	\left[
	\begin{array}{cc}
	\dot G(t) & G(t)\\
	\ddot G(t) & \dot G(t)
	\end{array}
	\right].
\end{equation*}	
The Green's function $G$ can not be evaluated in general
in a closed form, the case of Ohmic damping being an exception,
where the spectral density is for small frequencies linear
in the frequencies. The Laplace transform $\hat G$
of $G$ is related to the Laplace transform $\hat\gamma$
of $\gamma$ 
as $\hat G(z) =(z^2 + z \hat \gamma(z) )^{-1}$.
In order to specify the long time 
behavior of the Green's functions, 
it is sufficient to know the power law for the 
spectral density for small frequencies only. Using
Eq.\ (\ref{spectraldensity}), one arrives for
$ p \in(0,2) $ at
$\lim_{t\rightarrow \infty} G(t)/f(t)=1$
(see also Ref.\ \cite{BroMo}), 
where
	$f(t)=  \sin(\pi p/2)  t^{p-1}/( \zeta \Gamma(p))$.
From the asymptotic 
behaviour of $f$ as $t\rightarrow\infty$ 
it can be seen after a few steps that 
$
    \lim_{t\rightarrow\infty} A(t)/A'(t)=1$,
with 
\begin{eqnarray*}
    A'(t)=\int_{-\infty}^t ds \int_{\infty}^t ds'
	G(t-s)
	G(t-s') K(s-s'),
\end{eqnarray*}
This quantity in turn 
happens to be a quantity 
investigated in  Ref.\ \cite{BroMo}, where it has
been shown that 
\begin{equation*}
    \lim_{z\rightarrow 0}
   (\hat A'(z)\beta z)/(2 \hat G(z))=1, 
\end{equation*}
which yields $\lim_{t\rightarrow\infty} 
A(t)/ A''(t)=1$,
with 
\begin{eqnarray}\label{a}
    A''(t)& =&
    \frac{2\sin(\pi p/2)}{
    \beta\zeta \Gamma(p+1)}
	t^{p}.
\end{eqnarray}
In order to find the long-time behaviour
of the function $C$, we may use the fact that
   $ C(t) = 2 \dot A(t)$
for all $t\in [0,\infty)$, which holds since
$G(0)=0$, and apply l'Hospital's rule to 
arrive at
$\lim_{t\rightarrow\infty} 
C(t)/ C''(t)=1$,
with 
\begin{eqnarray}\label{b}
    C''(t)& =&
    \frac{2p \sin(\pi p/2)}{
    \beta\zeta \Gamma(p+1)}
	t^{p-1}.
\end{eqnarray}
To get the long term behaviour of $B$, we
can again start with
that $
    \lim_{t\rightarrow\infty} B(t)/B'(t)=1$,
where
\begin{equation*}
    B'(t)=\int_{-\infty}^t ds \int_{\infty}^t ds'
	\dot G(t-s)
	\dot G(t-s') K(s-s').
\end{equation*}
This, in turn, is nothing but the momentum
uncertainty in the stationary
setting, which is well-defined even in this
free case (compare also Ref.\ \cite{Ford,Scully}),
\begin{equation}\label{c}
    \frac{1}{\pi}
    \int_{0}^{\infty}d\omega
    \text{im}[\alpha(\omega+ i 0^+)]   \omega^2 \coth(\omega\beta)= B_{\infty}>0,
\end{equation}
with $\alpha(z)=-1/(z^2 + i z \tilde \gamma(z))$.
$B_\infty$ 
is a (time-independent) positive 
real number. 
So we have determined the long-time behaviour
of the entries of the symmetric
$2\times2$-matrix $M(t)$.

Subsequently, a pointer state is taken to be
a minimal uncertainty Gaussian state with
particular second moments that reflect a small 
uncertainty in the position canonical coordinate.
The statements will be formulated in a language common in 
quantum optics and continuous-variable quantum information
theory. The first moments
are
    $(d_{1},d_{2})= (\langle X \rangle,
    \langle P \rangle)$,
the second moments are collected in the
covariance matrix
\begin{equation*}
    \Gamma=  \left[
    \begin{array}{cc}
	2 \langle O_{1}^{2}\rangle & 
	\langle O_{1}O_{2}+ O_{2}O_{1}\rangle \\
	\langle  O_{1}O_{2}+ O_{2}O_{1}\rangle & 
	2\langle O_{2}^{2}\rangle
	\\
    \end{array}	
    \right],
\end{equation*}
where 
$O_{1}=X-\langle X\rangle$ and
$O_{2}=P-\langle P\rangle$.
The second moments for the pointer states
are taken to be
\begin{equation}\label{sm}
    \Gamma_{\infty}=\left[
    \begin{array}{cc}
    B_{\infty}^{-1}& 0\\
    0 & B_{\infty}\\
    \end{array}
    \right].
\end{equation}
This is a covariance matrix of a minimal 
uncertainty state, as $|\Gamma_{\infty}|=1$.
Note that in
the weak damping limit, 
$B_{\infty}$
becomes approximately \cite{Sense}
   $ B_{\infty}= \beta^{-1}=T$
(note that $\hbar=1$ and $k=1$),
so in the weak coupling and high temperature limit,
the set of pointer states is a set of minimal
uncertainty Gaussian states very narrow in position.
The corresponding pure Gaussian 
state with first moments
$(d_{1},d_{2})=(\xi_{1},\xi_{2})$
will be denoted as
\begin{equation*}
    \rho_{\xi}=|\psi_{\xi }\rangle
    \langle \psi_{\xi }|.
\end{equation*}
This set of minimal uncertainty Gaussian states, which becomes a set of 
states very narrow in position in the limit of weak coupling and high 
temperatures, will be regarded as the set of pointer states \cite{Amb}.
It is an overcomplete set of states satisfying
    $|\langle\psi_{\xi}|\psi_{\xi'}\rangle |^{2}=
    e^{- (\xi-\xi')^{T}( \Gamma_{\infty}/2) (\xi-\xi')}$.
The analogue of the standard $s$-ordered
Wigner function of a state $\rho$
may be defined as
\begin{eqnarray*}
    W_{s}(\xi) &=& \frac{1}{\pi^{2}}\int d^{2}\xi'
    e^{s ({\xi'_{1}}^{2} + {\xi'_{2}}^{2})/ \sqrt{2}}
    e^{-2i \xi\sigma {\xi'}^{T}}\nonumber \\
    &\times &
    \text{tr}[e^{i\alpha \{X,P\}_{+}  }
    e^{ i (\xi_{1}X+ \xi_{2}P)} e^{-i\alpha\{X,P\}_{+}  } 
    \rho  ], 
\end{eqnarray*}
$s\in[-1,1]$,
where $\sigma$ is the symplectic matrix embodying the canonical commutation 
relations, $\{.,.\}_{+}$ denotes the anticommutator, and 
 $   \alpha = -\log(B_{\infty})/2$
is the squeezing parameter corresponding to the pointer states 
(taken with respect to the standard unit quantum oscillator). 
The state can then 
be represented as \cite{Schleich}
\begin{equation}\label{re}
    \rho=\int d^{2}\xi W_{1}(\xi) 
    |\psi_{\xi}\rangle\langle\psi_{\xi}|,
\end{equation}
whereas in turn
$
W_{-1}(\xi)=\langle \psi_{\xi}| \rho |\psi_{\xi}\rangle/\pi\geq 0
$
for all $\xi\in \rr^{2}$.
Then, the $s$-ordered functions are related to each other
via convolutions
(compare, e.g., Ref.\ \cite{Max})
\begin{eqnarray}\label{co2}
    W_{s}(\xi) =\int d^2{\xi}'
    \frac{W_{s'}(\xi')}{2\pi}
    \frac{4}{s'-s}
    e^{- 2 (\xi-\xi') \Gamma^{-1}_{\infty} 
    (\xi-\xi')^{T}/(s'-s)},
\end{eqnarray}
for $s<s'$. We are now in the situation that we can 
argue similarly to 
Ref.\ \cite{Kiefer}: the function $W_{0}':\rr^{2}\times 
\rr^{+}\longrightarrow \rr$,
   $W_{0}'(\xi,t)=
    W_{0}(V^{-1}(t)\xi,0)/|V(t)|$,
is a legitimate Wigner function, as can be read off the definition
of the Wigner function. 
Then,
Eq.\ (\ref{solution2})
and Eq.\ (\ref{co2}) imply that 
\begin{eqnarray*}
    W_{1}(\xi,t) &= & 
   \int d^2{\xi}'
    \frac{W_{0}'(\xi',t )}{2\pi}
    |M(t) - \Gamma_{\infty}/4
    |^{-1/2}\nonumber\\
    &\times&
    e^{- (\xi-\xi')(M(t) - \Gamma_{\infty}/4)^{-1} (\xi-\xi')^{T}/2}.
\end{eqnarray*}
But since
\begin{eqnarray*}
    \int d^2{\xi}'
    \frac{W_{0}'(\xi',t)}{2\pi}
    4  
    e^{- 2 (\xi-\xi') \Gamma_{\infty}  (\xi-\xi')^{T}}\geq0
\end{eqnarray*}
for all $\xi\in \rr^{2}$, $W_{1}(\xi,t)\geq 0$ for all $\xi\in 
\rr^{2}$ if
\begin{eqnarray}\label{fin}
    M(t)- \Gamma_{\infty}/4\geq \Gamma_{\infty}/4.
\end{eqnarray}
In turn, given the time dependence of the coefficients of $M(t)$, 
demonstrated in Eqs.\ (\ref{a}), (\ref{b}), and (\ref{c}),
there exists a finite $t_{c}>0$ such that (\ref{fin}) is valid for all $t>t_{c}$.
This time $t_{c}$, in turn, is the time from which on $W_1$ is strictly positive, and 
the state can certainly 
exactly
be represented as a mixture of pointer states with
second moments as in Eq.\ (\ref{sm}). 

\begin{figure}

\includegraphics[width=5.0cm]{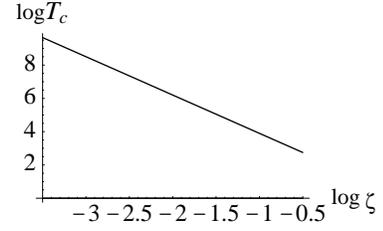}

    \caption{This figure shows $\log T_c=\lim_{T\rightarrow\infty} \log t_{c}$, where
    for a given temperature the number $t_{c}>0$ is the smallest number 
    such that (\ref{fin}) is satisfied
    for all $t>t_{c}$  
    for the case of strictly Ohmic damping, as a function of $\log\zeta$. 
    The stronger the damping, 
    the faster is the 'localization' process.}
    \label{figure}
\end{figure}

This is a generic result for arbitrary non-zero temperatures,
arbitrary coupling strengths and all the spectral densities
as in Eq.\ (\ref{spectraldensity}). 
For specific choices for
the spectral density, bounds for the
time $t_{c}$ can be found from which 
on the state can be represented
as a mixture of pointer states.
For Ohmic damping in particular, 
the Green's function is given by 
$    G(t)= 
    (
    1-e^{-\zeta t}
    )/\zeta$, i.e., $\hat \gamma(z)=\zeta >0$ \cite{BroMo,Ford}.
The behaviour 
becomes particularly transparent 
in the high temperature case. We then simply obtain
$	\lim_{T\rightarrow\infty}A(t)/T =   2(t-G(t))/\zeta$, 
$	\lim_{T\rightarrow\infty} C(t) /T  = 2(1-\dot G (t))/\zeta$, and
$	\lim_{T\rightarrow\infty} B(t)/T  =    1-e^{-2\zeta t}$.
Fig.\ \ref{figure} depicts $ T_c=\lim_{T\rightarrow\infty}t_{c}$, where
$t_c$ is the smallest  time
 for which (\ref{fin}) is satisfied for strictly 
Ohmic damping.

To conclude, it has been shown  
that if one couples
a free particle linearly to a heat bath prepared in the Gibbs state
of some temperature, then, under very general conditions and without 
approximations, 
the state of the system becomes after some finite 
time exactly indistinguishable from an exact
mixture of particular
minimal uncertainty Gaussian pointer states. In this sense
it can be said that exact decoherence to these localized pointer
states is generic, and not only a 
feature of a limit that can be regarded as being classical. Locally,
hence, we arrive at the situation as if we had merely classical 
ignorance about the position of the particle. 
Needless to say, care is required
in the interpretation of the result, and one should not be
tempted by a realistic interpretation in terms of classical
alternatives. In turn, 
the total state of both the system and its environment is very 
different in structure and is typically a highly correlated and 
often, but not necessarily  \cite{Eisert}, 
entangled state. It is the hope that 
this paper can
contribute to the debate on 
the dynamical appearance of classical properties in 
quantum theory. This debate is potentially 
becoming more timely than ever with
the availability of novel experiments on decoherence  \cite{Exp}, 
let it be with microwave cavities, ion traps, 
or nano-electromechanical systems.

I would like to thank 
W.H.\ Zurek,
H.-P.\ Breuer, M.\ Cramer, J.J.\ Halliwell 
and C.\ Henkel
for very thoughtful and detailed
comments on the manuscript and M.B.\ Plenio and P.\ H{\"a}nggi 
for discussions.
This work was supported by the EU 
(QUPRODIS, QUIPROCONE), 
the A.v.-Humboldt-Foundation, the ESF, and the DFG.


\begin{thebibliography}{99}
\bibitem{NoteFact}
	Note that it is an interesting problem in its own right to
	investigate the influence of correlated initial conditions, such that
	time evolution of the state of the distinguished system no longer 
	corresponds to a completely positive map. This is
	relevant, e.g., when considering a charged quantum particle coupled 
	to the 
	radiation field, where the assumption of 
	product initial conditions is typically not invariant under
	gauge transformations \cite{QED}.
\bibitem{Wigner}
	The convention for the Wigner function is 
	\begin{eqnarray*}
	W(\xi) &=&\frac{1}{\pi^{2}}\int d^{2}\xi'
	e^{-2i \xi\sigma {\xi'}^{T}}
	\text{tr}[e^{ i (\xi_{1}X+ \xi_{2}P)}\rho],\,
		\sigma =\left[
	\begin{array}{cc}
	0& 1\\
	-1 & 0
	\end{array}
	\right],
	\end{eqnarray*}
	with 
	$\sigma$
	being the symplectic matrix, 
	such that $\xi\sigma {\xi'}^{T}$ is the standard
	symplectic
	scalar product between $\xi, \xi'\in \rr^{2}$.
\bibitem{Units}	
	For simplicity of the argument, the 
	mass of the particle will set to unity and
	units will be employed where
	$\hbar=1$ and $k=1$.
\bibitem{Sense}
	This is meant in the sense that for any $\varepsilon>0$
	there exists a $\zeta>0$ such that $|B_{\infty}-T|<\varepsilon$.
\bibitem{Amb}
	Needless to say, there is a certain degree of ambiguity in the choice 
	of the pointer states here. The key argument is that the final state 
	can after a  finite time  be exactly represented as a convex combination of 
	these reasonable pointer states.
\end{thebibliography}
\end{document}